\documentclass{article}

\begin{document}

\title{Different approaches to describe
depletion regions in first order phase transition kinetics}

\author{Victor Kurasov}

\date{Victor.Kurasov@pobox.spbu.ru}

\maketitle

During the first order phase transitions the
objects of a new phase consume a mother phase
which leads to some density gaps near existing
embryos of a new phase. This fact is rather
obvious \cite{stowell}, but complete theoretical
description was given only in \cite{kurasov}. Now
this fact in widely recognized  \cite{korsakov}
but still there are some attempts \cite{zuv} to reconsider
this approach which is now   known as the model of
depletion zones \cite{korsakov}. Here we shall
analyze \cite{zuv} and show the useless of
reconsideration proposed in \cite{zuv}.

We don't consider here the picture of the gobal
evolution of a system - in \cite{zuv} it is so
primitive that can not give any reliable
quantitative results but only some very rough
estimates without any justifications.
We needn't to do it because this has been already
done in \cite{kurasov} with all necessary
justifications. So, we consider only the profile
around the solitary embryo,  which was the matter of interest in 
\cite{zuv}. Why it is possible to
consider only the solitary embryo? An answer has
been given in \cite{kurasov}, in \cite{zuv} this
question remains unsolved. Due to results of
\cite{kurasov} we shall consider namely the
solitary embryo.

We assume that the substance exchange regime is a
diffusion one. Then there is a gap of a vapor
density $n$ near an embryo. Then the density $n$
differs from a density $n (\infty)$. This
difference initiates a variation in a nucleation
rate $I$ and causes the effect of depletion
zones. So, it is very important ro get for $n(r)
- n(\infty)$ ($r$ is a distance from a center of an embryo)
a correct expression. This
expression is a crucial point in consideration of
an effect of depletion zones in nucleation
kinetics. Certainly, it has to be a function of
time $t$.

In \cite{zuv} the following expression is
proposed
\begin{equation} \label{1}
n(r,t) = n_s(r) +
(n(\infty) + n(0)) \frac{R}{r} \Phi (\frac{r-R}{
2 \sqrt{Dt}})
\end{equation}
\begin{equation} \label{2}
n_s (r) = n(\infty) -\frac{n(\infty) - n(0)}{r}R
\end{equation}
Here $n(0)$ is equilibrium concentration on the surface of nucleus,
$D$  is a diffusion coefficient, $R$ is the
nucleus radius, $\Phi(x) $ is the Laplace
probability integral.

Here the sign $+$ was used instead of $+$
in the second term of (\ref{1}).
Correct solution in the neigbourhood of an embryo
is \cite{eger}
\begin{equation} \label{3}
\frac{n-n(\infty)} { n(0) - n(\infty)} =
\frac{R}{r} erfc(\frac{r-R}{2\sqrt{Dt}} )
, \ \ \  erfc (z) = \frac{2}{\sqrt{\pi}}
\int_z^{\infty} \exp(-\xi^2) d\xi
\end{equation}

But even with this solution one can not construct
the correct form of the nucleation rate profile. We
shall outline this point.

There are following necessary suppositions to get
this exact solution for concentration profile:

\begin{itemize}
\item
The radius of an embryo is constant in time

\item
The boundary condition at $r=R$ is constant in
time.

\end{itemize}

Both assumptions are rather approximate. As a
result the above solution is valid only near the
surface of an embryo. It is easy to see from the
following arguments. Let $\delta t$ be the
characteristic time when the relative variation
of intensity of vapor
consumption $ v $ is already essential
$$
|v(t+\delta t ) - v(t)| \sim v(t)
$$
Then it follows that only  the distances
$$
r-R \sim 4 D \delta t \equiv \delta R
$$
or smaller ones
can be considered on the base of solution
(\ref{3}). At all  distances $r-R > \delta R$
we have to take into account that earlier the
intensity of vapor consumption was smaller.
Really
$$
v \sim d\nu / dt
$$
where
$\nu$ - a number of molecules inside the embryo.
As it is known
$$
d \nu / dt   = const  \  \nu^{1/3}  \sim t^{1/2}
$$
and we see that the intensity isn't constant in
time.

It seems that the small distances are the main
ones. Certainly the gap is greater the smaller
the distance is. But the small distances aren't
essential because the functional form of the
nucleation rate \cite{kurasov} can be
approximated as
\begin{equation}\label{i}
I(r) =
I(r=\infty) \exp(\Gamma ( n(r)-n(\infty) ) /n(\infty))
\end{equation}
where $\Gamma$ is a big parameter. It is
approximately equal to $\nu$.
Note, that the sence of nucleation rate as a
probability to appear for an embryo was used here
(details see in \cite{kurasov}).
So, the distances where
$$
-
\Gamma ( n(r)-n(\infty) )  /n(\infty) \approx 1
$$
are the most interesting.
When
$$ |
\Gamma ( n(r)-n(\infty) ) /n(\infty) | \ll 1
$$
we have
$I=I(\infty)$ and the is no gap of concentration.
When
$$ -
\Gamma ( n(r)-n(\infty) ) /n(\infty) \gg 1
$$
we have
$I=0 $ and the is no nucleation. So, this region
isn't interesting.

The distances $
\Gamma ( n(r)-n(\infty) ) /n(\infty) \approx 1
$ because of $\Gamma \sim \nu \rightarrow
\infty$ corresponds to very small
$n(r)-n(\infty)$ and, thus, to big 
values of distances
$ \sim \delta r $ from the embryo.
Ordinary $\delta r \gg \delta R$ and there is no
possibility to use (\ref{3}).

Instead of (\ref{3}) in \cite{kurasov} another
approach was used. This approach is based on the
strong inequality
 \begin{equation}\label{m}
 \delta r \gg R
 \end{equation}

Why this inequality takes place? When we consider
classical nucleation, i.e. transition of a
supersaturated vapor into a liquid phase
everything is clear. We have a  strong inequality
$$
v_v / v_l \gg 1
$$
where $v_v$ is a partial volume for one molecule
in a vapor, $v_l$ is a partial molecule in a
embryo.
This inequality makes obvious that the final
volume after the whole process of condensation is
small. As an example of condensation of water
vapor in normal external conditions.
Suppose that supersaturation is
somewhere $\sim 5$. Then because $v_v/ v_l \sim
1000$ we see that the final volume 
of a new phase is very small
$\sim 1/200$ and the characteristic distance
$\delta r \sim (200)^{1/3} R \gg R$.

We see that the value $\delta r$ has to take the
same order of magnitude as the mean distance between
objects of a new phase
$r_{mean}$. Really, when $\delta r$ attains
imaginary value $r_{mean}$ it means that the
process of nucleation stops. So, namely the
values at this moment are the final ones and the
main ones. Then
$$
r_{mean} \sim \delta r
$$
The same
will be valid for all other first order transitions.

Consider the opposite transition from the liquid to
the
vapor phase. We shall see the opposite
inequality. But we know that compressibility of
liquid is very law. So, the relative super
streching is rather low. It means that the vapor
phase in a final state will occupy some rather
small volume. And again we see that the mean
distance between the objects of a new phase
is many times greater than the size of objects.
Again we get the same result.

We see that the following fact takes place:
{\it
The relative final volume of a new phase will be
small.
}

To see that (\ref{m}) takes place we needn't to
consider the real final volume of a new phase
after phase transition but can take the values at
the end of nucleation. Because the relative quantity of
surplus
substance condensed in a new phase has an order
$\Gamma^{-1}$  (at least under the collective
regime of substance consumption \cite{kurasov})
we see that the relative volume occupied by a new
phase is limited from above by a value of an
order $\Gamma^{-1}$.  So
$$
r_{mean} \sim R \Gamma^{1/3}$$
and the required property is established.

So, we see the required property for moderate
effects of depleted zones. When the effect of
depletion zones is strong we can use at $r \sim (2
\div 3 ) R $ stationary solution and see that
$\delta r > (5\div 7)R $. So, here the required
property is also observed which completes
justification.

Then we can use profiles obtained on the
formalism of Green function. We have
\cite{kurasov}
\begin{equation}\label{g}
n(\infty) - n(r) =
\int_0^t \frac{\lambda x^{1/2}}{8 (D \pi (t-x))^{3/2}}
\exp(-\frac{r^2}{4D(t-x)}) dx
\ \
\end{equation}
\begin{equation}\label{g1}
\lambda =  (4 \pi)^{3/2}
 (\frac{v_l}{2\pi})^{1/2} (\zeta n_{\infty} D )^{3/2}
\end{equation}
and $\zeta$ is a supersaturation.

This result was obtained for transition from
vapor phase into a liquid phase but there is no
difference because the kinetic mechanismremains
the same.

One can see  that expressions (\ref{1}),
(\ref{2}) and (\ref{g}), (\ref{g1}) are
absolutely different
in analytical structure. The same is valid when we
compare (\ref{1}),
(\ref{2}) and (\ref{g}), (\ref{g1}) numerically.

On the base of profile we can easily get all
results of the nucleation process.

The rate of nucleation can be found according to
(\ref{i}). Then we have
\begin{equation}\label{i2}
I(r) =
I(r=\infty) \exp(\Gamma
{ n(0) - n(\infty)}
\frac{R}{r} erfc(\frac{r-R}{2\sqrt{Dt}} )
 /n(\infty))
\end{equation}
for analytical solution with fixed boundary 
of embryo and 
\begin{equation}\label{i3}
I(r) =
I(r=\infty) \exp(-\Gamma
\int_0^t \frac{\lambda x^{1/2}}{8 (D \pi (t-x))^{3/2}}
\exp(-\frac{r^2}{4D(t-x)}) dx
/n(\infty))
\end{equation}
for solution on the base of Green functions.
Note that the analogous transformation in
\cite{zuv} was done with partial shift of
coordinate $r-R$ to $r$ which is wrong and 
beside this isn't
necessary.

We see that these approaches give absolutely
different results.

The next step is an obvious remark that the
probability $dp$ that an embryo appears during
elementary interval $dt$ at the distance from $r$
up to $r+dr$ from the center of the already
appeared embryo is
$$
dp = dt 4 \pi r^2 I(r,t)
$$
Then the probability $dP$ that the embryo appears
in the layer from $R$ up to $r$ is
\begin{equation}\label{dp}
dP = dt \int_R^r 4 \pi r'^2 I(r',t) dr'
\end{equation}

This expression differs from \cite{zuv} where the
following expression was presented
$$
dP = dt \int_r^R 4 \pi r^2 I(r',t) dr'
$$
The last expression is wrong and we shall follow
(\ref{dp}).

The next step used in \cite{zuv} is to
come to an integral value $P$ which is the
probability that in a sphere of radius $r$ around
already existing embryo there will be no
appearance of a new embryo. Certainly we can
start from  (\ref{dp}) and integrate it.  But it
is necessarfy to take into account that the zone
of depletion will grow according to the absolutely
precise law. This law was established in
\cite{kurasov}. But here we 
suppose\footnote{To
be close to  \cite{zuv}.} 
that one can act in another manner
and write
\begin{equation} \label{e}
P = \exp(- \int_0^t \int_R^r 4 \pi r'^2 I(r',t') dr' dt')
\end{equation}

The last expression analogous to those
which forms the base for
further constructions
in\footnote{In \cite{zuv} an
intergal over $r$ is  from $0$ up to $r$.
} \cite{zuv} is rather
doubtful.  The problem appears because now we are
coming to characteristic of a whole process of
nucleation (earlier there was a profile around a
solitary object of a new phase).

For a "solitary
subsystem" the derivation of (\ref{e}) is evident:
Let
$$
dp = dt I
$$
 be a probability that during an
elementary interval $dt$ there will be appearance
of a new embryo.  The value of $dt$ is rather
small. Then $dp$ is proportional to $dt$ and to
the rate of nucleation $I$ (here the sense of the
nucletaion rate as a probability is used).
Then the probability for the absence of
appearance is
$$
dp' = 1-dp = 1-dtI
$$
If we have two elementary intervals $dt_1$ and
$dt_2$ which don't overlap then the probability
of the absence of embryo appearance is
$$
dp_1' dp_2' = (1-I_1dt_1)(1-I_2 dt_2)
$$
The same is valid for arbitrary number of
intervals.

To fulfill
multiplication in the r.h.s. one can present
$dp'$
as
$$
dp' = (1-I dt) = \exp(-I dt)
$$
Then the total probabiblity of the absence of
appearance is
\begin{equation}\label{ee}
P = \prod_i dp_i' = \exp(-\sum_i dt_i I_i) =
\exp(-\int I(t) dt)
\end{equation}
and we come to (\ref{e}).

The real problem is how to take into account the
overlapping of density profiles initiated by
different embryos. This problem was solved in
\cite{kurasov}. The exponential form (\ref{e})
for $P$ is 
some rather arbitrary  
interpretation of the value "the
free volume" used in \cite{kurasov}. Now we shall
discuss this interpretation in frames 
of the approximation  of solitary droplet.

The overlapping of exhausted regions (ER)
requires to use instead of (\ref{i1}) -
(\ref{i3})another approximation where instead of
one profile there is a superposition of profiles.
Certainly it is impossible to calculate this
superposition precisely. Then one will come to
some models analogous to formulated in
\cite{kurasov}.

One can directly use the results from
\cite{kurasov} here.  When the free volume
\cite{kurasov} is calculated as function of time
one can determine the total nucleation rate as a
ratio of the free volume to the whole volume of a
system and use then (\ref{ee}).

The real problem is what consequences can we make
from a knowledge of $P$.
On one hand it seems that the knowledge of $P$
solves al problems in knetics of the global
nucleation stage. Really, having presented $P$
in a form
$$
P = \exp(-L(r,t))
$$
where $L(r,t)$ is some expression one can
approximately estimate the average radius of 
a sphere 
where
there will be no appearance by relation
$$
L(r,t \rightarrow \infty) |_{r=r_0}= 1
$$
This construction analogous to \cite{zuv}
needs two remarks.

This estimate makes no difference between embryos
appeared earlier or later during the nucleation
stage.  It means that this approximation
supposes that the spectrum of droplets sizes is
monodisperce one. It isn't correct assumption and
can be treated only as a rough estimate.

It is known that sometimes the overlapping of
profiles is important and makes the main
contribution in nucleation kinetics.  This
phenomena can occur when the substance exchange
regime is goimg to be a free molecular one and
when the  long tails of profiles are important.
The  first situation can not be realized here.
The last situation was a metter of separate
consideration in \cite{kurasov2}.

In  \cite{kurasov} an integral definition of a
boundary of ER allowed to take into account the
situation of long tails. Here this situation can
not be considered, because the level type  definition
is used. It means that the value of $r_0$ is
determines as a value when $L$ attains some level
($=1$). The definition of level type can not take
into account long tails and it is more convenient
to use approach from \cite{kurasov}. Here this
approach is considered because it is analogous
to \cite{zuv}.

When the value of $r_0$ is determined it is easy
to estimate the total average number of droplets $N$
as
$$
N= \frac{3}
{4\pi} r_0^{-3}
$$
(the coefficient $\frac{3}
{4\pi} $ can be omitted, this 
depends on interpretation of $r_0$).

To fulfill concrete calculations one can use the
following  approach. Note that $P$ can be in any
case presented as
$$
P = \exp(- \int_0^{\infty} dt' \int_{r_l}^r dr'
4 \pi r'^2 \exp(f(r',t'))
$$
The lower limit $r_l$ of integration has to be put 
$r_l = R$
 for  models with
explicit boundary of an embryo and for models with
 Green function $r_l = 0$.
But because $\delta r \gg R$ the effect of the 
lower limit will be small and we can put $r_l = 0$ in
ll situations.  Certainly, for $P$ due to the exponential form the 
effect is essential,  but  when we are calculating the mean 
distances 
the essential dependence on $r_l$ disappears due to 
$\delta r \gg R$.
In the last equation $f$ is some function with
explicit form given by solution of diffusion
equation.

One can see that in $f(r,t)$ two variables has to
appear in combination $\beta = t/r^2$ or 
$\beta' = t/(r-R)^2$ . The
dependence on $t$ and $r$ via $\beta$ is the main
one (in special regimes of mother phase
consumption there may be dependence on $t,r$ in
another combination, but this dependence will be
much more smooth than via $\beta$).
Because $\delta R \gg R$ the difference in substitution 
$\beta'$ 
instead of $\beta$ 
will be small (not in $P$, but on mean distances 
and times).

Then in integration $\int_0^t dt'$  we have to
substitute $dt'$ by  $ d \beta' $ which gives
$$
P = \exp(-  \int_0^r dr'
4 \pi r'^4 const
$$
where $const$ comes from
$$
const =  \int_0^{\infty} d\beta'
 \beta'^2 \exp(f(r',t'))
$$
because
$$
\int_0^{\infty} dt' = r'^2 \int_0^{\infty}
d\beta'
$$

It leads to
\begin{equation}\label{rep}
P = \exp(-r^5 \gamma)
\end{equation}
where $\gamma= 4\pi const/5$ is some constant.

One can show that
$$
\int_0^{\infty} dx \exp(-x^5)   \approx 1
$$
This approximate equality is very important.
Namely this equality allows to determine the mean
distance between droplets as $N^{-1/3}$, or as
$r_0$ or according to the sense of $P$. 
Three ways are available. They have to give similar results.
This fact is ensured by this approximate
coincidence. Then the third way gives
$$
\bar{r} = \int_0^{\infty} d r P r =
\frac{C_0}{2 \gamma^{2/5}} \int_0^{\infty} d
\alpha \exp(-\alpha^{5/2})
$$
where $C_0$ is the normalizing factor of
distribution $P$. One can calculate $C_0$
according to
$$
C_0 \int_0^{\infty} dr \exp(-\gamma r^5) =1
$$
and
$$
C_0 = \frac{\gamma^{1/5}}{\int_0^{\infty} dx
\exp(-x^5)}
$$
Then\footnote{This value is calculated in \cite{zuv}
in a wrong way.}
$$
<r> =
\frac{1}{2 \gamma^{1/5}}
\frac{\int_0^{\infty} d
\alpha \exp(-\alpha^{5/2})}{\int_0^{\infty} dx
\exp(-x^5)}
$$
Because
$$
\int_0^{\infty} d
\alpha \exp(-\alpha^{5/2})
\approx
1
$$
$$
\int_0^{\infty} dx
\exp(-x^5)
\approx 1
$$
we see that
$\bar{r}$ is two times smaller than the mean
distance between embryos. It is clear because the
mean distance between embryos is two mean
distances until the boundary of exhaustion zone
$r_0$. So, we observe a coincidence. Certainly,
one can calculate integrals more precisely.

The average time of waiting fro the appearance of
new embryo near the already existing one is
directly and elementary 
connected with a mean distance between
embryos and we needn't to calculate it
separately\footnote{The integral in expression for
this value is calculated in  \cite{zuv} in a wrong way}.

As for calculation
a quadratic mean deviation from the average
distance it is given\footnote{Result in \cite{zuv}
is wrong.} by
$$
<(r-<r>)^2> =
<r>^2 (\frac{I_2I_0}{I_1^2}- 1)
$$
where
$$
I_i = \int_0^{\infty} dx x^i \exp(-x^5)
$$
But we have to stress that this result has
practically no meaning because as it is stated in
\cite{stoh} the effect of interaction leads to
 decrease of fluctuation of the total
number of droplets.  The right numerical value
of this effect is given in \cite{kurstoh}.
Here we have absolutely no interaction between
droplets and this result has to be seriously
changed due to interaction. An example of such
account is given in \cite{PhysRev2001}.

To close this question we shall
 construct  now more precise solutions
of diffusion equation.
They have to be used in (\ref{i}) and then in (\ref{e})

We shall start with the following model.
There exists a more general solution
\cite{eger}.

For a domain $R<r<\infty$ where
$$
n = f(r) \ \ at \ \ t=0
$$
and
$$
n = g(t) \ \ at \ \ r=R
$$
the solution of diffusion equation
$$
\frac{\partial n }{\partial t} =
D
(
\frac{\partial^2 n }{ \partial r^2} +
\frac{2}{r} \frac{\partial n}{\partial r}
)
$$
(this the diffusion equation with spherical
simmetry)
is \cite{eger}
\begin{eqnarray}
n = \frac{1}{2r\sqrt{\pi Dt}} \int_R^{\infty} \xi
f(\xi) [\exp(-\frac{(r-\xi)^2}{4Dt}) -
\exp( - \frac{(r+\xi - 2 R)^2}{4 D t} ) ]
d\xi
\nonumber
\\
\\
\nonumber
+
\frac{2R}{r\sqrt{\pi}}
\int_z^{\infty} g (t-\frac{(r-R)^2}{4 Dt} )
\exp(-\tau^2 ) d \tau
\end{eqnarray}
where
$$
z= \frac{r-R}{2 \sqrt{Dt}}
$$

This solution allows to formulate the following
approximation.
Let $R$ be constant
$$
R=R(t=0)
$$
 but the boundary conditions
will be recalculated according to the stationary
solution:
\begin{equation}
n_s (R(t=0)) = n(\infty) -\frac{n(\infty) -
n(0)}{R(t=0)}R(t)
\end{equation}
Really the rate of embryo growth is so small
that the stationary distribution can be regarded
as practically precise one. So, this
approximation is practically precise and takes
into acound the variation of mother phase
consumption by an embryo.

Note that diffusion equation
$$
\frac{\partial w }{\partial t}
= D [ \frac{\partial^2 w }{\partial r^2}
+\frac{2}{r} \frac{\partial w}{\partial r} ]
$$
in the system with  spherical symmetry can be
transformed by substitution
$$
u(r,t) = r w(r,t)
$$
to an ordinary one dimensional diffusion equation
$$
\frac{\partial u}{\partial t} =
D \frac{\partial^2 u }{\partial r^2}
$$
We shall consider this equation in future.

Then we can make a shift  $x=r-R(0)$.
The problem
$$
w= f(x)
$$
at $ t=0$
$$
w=g(t)
$$
at $x=0$
has
solution
\begin{eqnarray}
w(x,t) = \frac{1}{2\sqrt{\pi D t}}
\int_0^{\infty}
[\exp(-\frac{(x-\xi)^2}{4 D t} ) -
\exp(-\frac{(x+\xi)^2}{4 D t} ) 
] f(\xi) d\xi
\nonumber
\\
\\ \nonumber
+
\frac{x}{2\sqrt{D \pi} }
\int_0^t
\exp(-\frac{x^2}{4D(t-\tau)})
\frac{g(\tau)}{(t-\tau)^{3/2}}
d\tau
\end{eqnarray}

It is also possible to
go  to variable
$$
\rho = r - R(t)
$$
where $R(t)=  A t^{1/2}$ is a boundary of an
embryo.
Then in the diffusion equation beside 
$$
\hat{Q} u \equiv
\frac{1}{D} \frac{\partial u}{\partial t} -
 \frac{\partial^2 u}{\partial \rho^2}
$$
there appear some additional terms.
Having considered $\hat{Q}$ as a main operator 
 we come to an iteration
procedure where all other  terms  are assumed to
be known in a previous approximation (in zero
approximation they are 
omitted). So, at every step of iterations 
we have to
solve a problem
$$
w= f(x)
$$
at $ t=0$
$$
w=g(t)
$$
at $\rho=R(0)$
for equation
$$
\frac{1}{D} \frac{\partial u}{\partial t} =
 \frac{\partial^2 u}{\partial \rho^2} +
 \Phi' (\rho, t)
$$
may be with renormalized $D$. Here $\Phi'$ is
a known (at the previous step) function.
After the mentioned shift $x=\rho - R(0)$ we came
to a problem
$$
w= f(x)
$$
at $ t=0$
$$
w=g(t)
$$
at $\rho=R(0)$
for equation
$$
\frac{1}{D} \frac{\partial u}{\partial t} =
 \frac{\partial^2 u}{\partial x^2} +
 \Phi (x, t)
$$
with known $\Phi$.
The solution of the last problem is 
\cite{eger}
\begin{eqnarray}
u = \int_0^{\infty} G(x,\xi,t) f(\xi) d\xi+
\nonumber \\
 \frac{x}{2\sqrt{D\pi}}
\int_0^t \exp[-\frac{x^2}{4D(t-\tau)}]
\frac{g(\tau)}{(t-\tau)^{3/2}} d\tau
\\
\nonumber
+ \int_0^t \int_0^{\infty}
G(x,\xi,t-\tau) \Phi(\xi,\tau) d\xi d\tau
\end{eqnarray}
where
$$
G(x,\xi,t) =
\frac{1}{2\sqrt{\pi D t} }
[
\exp(-\frac{(x-\xi)^2}{4Dt})
-
\exp(-\frac{(x+\xi)^2}{4Dt})
]
$$
The last relation solves the problem to construct
iterations on every step.
Thus, we come to solution of diffusion equation
which has to be used in our previous
constructions.

\end{document}